\documentclass[twocolumn,aps,superscriptaddress,preprintnumbers]{revtex4}

\usepackage{amsmath}
\usepackage{graphicx}
\usepackage{amssymb}
\usepackage{epstopdf}
\usepackage{braket}

\usepackage{bm} 
\usepackage{longtable}
\usepackage{multirow}
\usepackage{enumerate}

\usepackage{chngcntr}

\usepackage{hyperref}
\hypersetup{
    colorlinks=true,       
    linkcolor=blue,        
    citecolor=blue,        
    filecolor=blue,        
    urlcolor=blue,         
    runcolor=blue
}

\usepackage{xcolor}

\begin{document}

\title{Vacuum-enhanced optical nonlinearities with organic molecular photoswitches}

\author{Marina Litinskaya}
\email{litinskaya@gmail.com}
\affiliation{Department of Physics \& Astronomy  and Department of Chemistry, University of British Columbia, Vancouver, Canada, V6T 1Z1}

\author{Felipe Herrera}
\email{felipe.herrera.u@usach.cl}
\affiliation{Department of Physics, Universidad de Santiago de Chile, Av. Ecuador 3493, Santiago, Chile.}
\affiliation{Millenium Institute for Research in Optics MIRO, Chile.}

\date{\today}           

\begin{abstract}
We propose a cavity QED scheme to enable cross-phase modulation between two arbitrarily weak classical fields in the optical domain, using organic molecular photoswitches as a disordered intracavity nonlinear medium. We show that a long-lived vibrational Raman coherence between the {\it cis} and {\it trans} isomer states of the photoswitch can be exploited to establish the phenomenon of vacuum-induced transparency (VIT) in high-quality microcavities. We exploit this result to derive an expression for the cross-phase modulation signal and demonstrate that it is possible to surpass the detection limit imposed by absorption losses, even in the presence of strong natural energetic and orientational disorder in the medium. Possible applications of the scheme include the development of organic nanophotonic devices for all-optical switching with low photon numbers. 
\end{abstract}

\maketitle

Organic chromophores and semiconducting polymers are known for having very large optical nonlinearities \cite{Chemla2012nonlinear}, with applications in lasing \cite{Kuehne2016}, frequency conversion \cite{Verbiest1997}, nonlinear microscopy \cite{Zipfel2003}, and all-optical switching \cite{Koos2009}. However, the typical magnitude of  the nonlinear susceptibility in organic materials is still orders of magnitude below those needed to enable nonlinear optical effects at the level of single photons, an important prerequisite in photonic quantum technologies \cite{OBrien:2009}. In recent years, the demonstration of strong and ultrastrong coupling of organic molecules with the quantized electromagnetic vacuum in optical and infrared cavities \cite{Lidzey1998,Ebbesen2016,Kena-Cohen2013,Long2015} has stimulated the study of single-photon control of chemical reactivity \cite{Hutchison:2012,Herrera2016,Galego2016}, energy transport \cite{Feist2015,Schachenmayer2015}, charge transport \cite{Orgiu2015,Pupillo2017}, and spectroscopy \cite{Herrera2017-PRL,Herrera2017-PRA,kowalewski2016cavity,Herrera2017-review}. However, the manipulation of polariton coherences to  enhance the natural optical nonlinearities of organic materials  still remains a largely unexplored topic \cite{Herrera2014,Liu:17,Barachati2018}.

We propose a scheme to observe cross-phase modulation between two {\it arbitrarily weak} classical fields within a nanoscale cavity, using disordered organic molecular photoswitches as the intracavity medium.  Molecular photoswitches are commonly used in photochemistry due to their ability to undergo  spectrally-resolved photoisomerization processes \cite{Dugave2003}, for applications such as optical information storage  and controlled catalysis \cite{Delaire2000,Kawata2000,Stoll2010}. 
In this work, we suggest that natural features in the vibrational structure of organic photoswitches can be exploited to enable the observation of vacuum-induced transparency (VIT \cite{Field1993,Rice1996,Tanji-Suzuki2011,Mucke2010}), a variation of electromagnetically-induced transparency (EIT \cite{Fleischhauer:2005}) in which a cavity vacuum--rather than a strong control laser-- is used to drive an internal coherence in the material.  {We further demonstrate that under conditions of VIT, the figure-of-merit for intracavity cross-phase modulation between the probe and signal fields can reach values that are orders of magnitude larger than without the cavity, exceeding the detection limit imposed by absorption losses.}  Replacing the control laser by a strongly coupled  vacuum  represents a clear advantage for organic materials with low laser damage thresholds \cite{Kuehne2016}. Although EIT has been measured at low temperatures in solid state systems with negligible inhomogeneous broadening \cite{Ham1997,Phillips2003,Wang2012,Acosta2013}, condensed-phase VIT has yet to be demonstrated in systems with strong inhomogeneously broadening, such as organic molecular ensembles. 
 
\begin{figure}[b]
\includegraphics[width=0.42\textwidth]{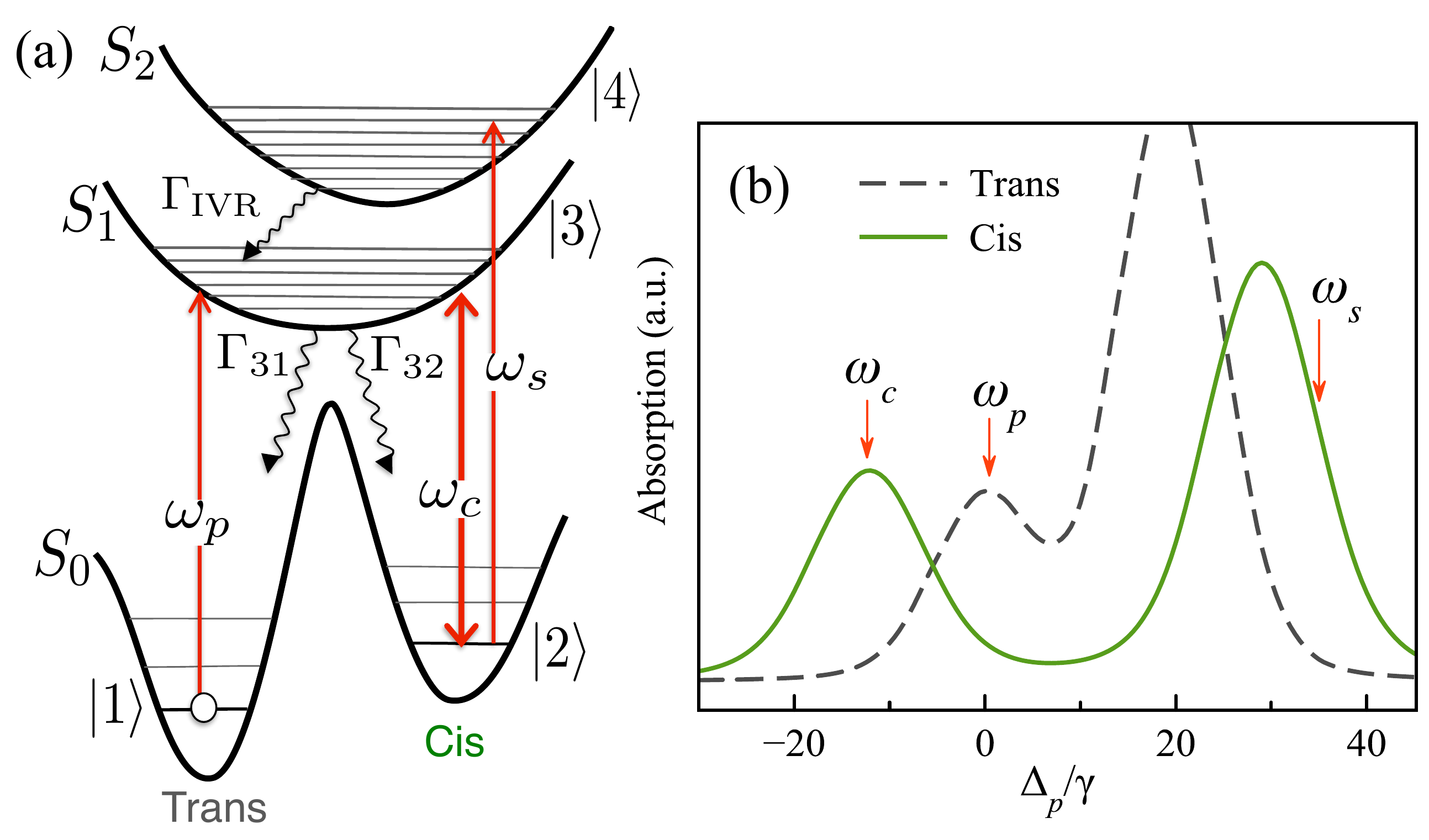}
\caption{Light-matter coupling scheme. (a) Energy level structure of molecular photoswitches. Straight arrows indicate near-resonant light-matter couplings at the cavity ($\omega_c$), probe ($\omega_p$) and signal ($\omega_s$) frequencies. Curly arrows indicate non-radiative molecular decay processes. (b) Representative absorption spectrum of the molecular photoswitch. The highest {\it trans} absorption peak corresponds to the transition $S_0\rightarrow S_2$.}
\label{fig:level-scheme}
\end{figure}

The envisioned light-matter scheme is shown in Fig.~\ref{fig:level-scheme}. The ground electronic potential of the photoswitch $(S_0)$ has two deep wells in the isomerization coordinate  corresponding to the {\it cis} and {\it trans} molecular isomers \cite{Dugave2003}. The lowest vibrational state of the {\it trans} isomer, is the stable ground state of the system $\ket{1}$. The meta-stable ground state $\ket{2}$ corresponds to the lowest vibrational level of the {\it cis} isomer potential well.  Thermal isomerization  of state $\ket{2}$ is strongly suppressed  \cite{Dugave2003}, resulting in lifetimes for state $\ket{2}$ of up to a few days at room-temperature \cite{Barrett1995,Dokic2009}.  We associate the lowest two excited electronic potentials of the molecular photoswitch, $S_1$ and $S_2$ \cite{Bandara2012,Tan2015}, with the states $\ket{3}$ and $\ket{4}$, respectively. The vibrational structure of the excited potentials is not resolved in condensed phase, but $S_0$ has well-defined torsional modes with frequencies in the range $\omega_{\rm v}\sim 160-190$ meV/$\hbar$ \cite{Armstrong1995,Hamm1997,Biswas2002}. The level structure in Fig. \ref{fig:level-scheme} generalizes the atomic  scheme introduced in Refs. \cite{Schmidt1996,Harris1998} to molecular systems with multiple ground state isomers.

Our model does not require an exact knowledge of the molecular potentials. Instead, we impose  a set of constraints on the inhomogeneously-broadened photoswitch absorption spectrum outside the cavity, which we illustrate in Fig.~\ref{fig:level-scheme}b. We assume that the photoswitch spectrum has: (\emph{i}) a well-resolved  band associated with the {\it cis} transition $\ket{2}\rightarrow \ket{3}$, resonant with the cavity frequency $\omega_c$; (\emph{ii}) a well-resolved band associated with the {\it trans} transition $\ket{1}\rightarrow\ket{3}$, near resonant with the probe frequency $\omega_p>\omega_c$; (\emph{iii}) a well-resolved high-frequency band associated with the {\it cis} transition $\ket{2}\rightarrow \ket{4}$, weakly driven by a blue-detuned signal field at frequency $\omega_s>\omega_p$. These spectral requirements can be met using  so-called {\it orthogonal} molecular photoswiches \cite{Bleger2012,Weston2014,Lerch2016}.

For a single molecular emitter, the light-matter coupling scheme in Fig. \ref{fig:level-scheme} can be modelled using the Hamiltonian 
\begin{eqnarray}\label{eq:Hamiltonian}
\lefteqn{\hat H_S=\omega_c\,\hat a^\dagger \hat a+ \omega_{21}\ket{2}\bra{2}+\omega_{31}\ket{3}\bra{3}+\omega_{41}\ket{4}\bra{4}}\nonumber\\
&&+\Omega_c(\ket{3}\bra{2}\hat a+\ket{2}\bra{3}\hat a^\dagger)+\Omega_p(\ket{3}\bra{1}{\rm e}^{-i\omega_p t}\\
&&+\ket{1}\bra{3}{\rm e}^{i\omega_p t}) +\Omega_s(\ket{4}\bra{2}{\rm e}^{-i\omega_s t}+\ket{2}\bra{4}{\rm e}^{i\omega_s t}),\nonumber
\end{eqnarray}
where $\Omega_p$ and $\Omega_s$ are the probe and signal semiclassical Rabi frequencies, respectively.  $\Omega_c$ is the cavity vacuum  Rabi frequency, and $\hat a$ is annihilation operator of the cavity field. In a dressed-state picture, our electronic basis must be supplemented with the cavity states to read: $\ket{\tilde 1}\equiv \ket{1;0_c}$, $\ket{\tilde 2}\equiv \ket{2;1_c}$,  $\ket{\tilde 3}\equiv \ket{3;0_c}$, and $\ket{\tilde 4}\equiv \ket{4;1_c}$. The probe field thus drives the material transition $\ket{\tilde 1}\leftrightarrow\ket{\tilde 3}$, the signal field the transition $\ket{\tilde 2}\leftrightarrow\ket{\tilde 4}$, and the cavity field strongly admixes the nearly-degenerate dressed states $\ket{\tilde 2}$ and $\ket{\tilde 3}$.  

We model energy disorder in the Hamiltonain $\hat H_S$ by defining the random transition frequencies $\omega_{31}=\langle \omega_{31}\rangle+\delta_{31}$, $\omega_{32}=\langle \omega_{32}\rangle+\delta_{32}$ and $\omega_{42}=\langle \omega_{42}\rangle+\delta_{42}$, where $\langle \omega_{ji}\rangle$ corresponds to the band center frequency and $\delta_{ji}$ is a random static fluctuation that gives rise to the inhomogeneous linewidths $\sigma_{ji}$.  Inhomogeneous linewidths due to energy disorder are typically in the range 150--200 meV for organic  photoswitches~\cite{Stuart2007},  magnitudes that far exceed a typical homogeneous linewidth. In addition to energy disorder, below we also model orientational disorder in the system by writing the vacuum Rabi frequency as $\Omega_c =\Omega_0\cos\theta$, where $\Omega_0$ is a constant amplitude and $\theta$ is a uniformly distributed random angle between the molecular transition dipole moment and the space-fixed cavity field polarization. 

Although in general the dissipative dynamics of photoisomerization involves a complex interplay between non-adiabatic  dynamics and non-secular relaxation \cite{Balzer2003,Rodriguez-Hernandez2014}, in this work we adopt a simpler Lindblad quantum master equation approach to describe dissipation. We account for empty cavity decay with  bare photon lifetimes assumed in the range $1/\kappa\sim 1-100$ ps, which can be achieved using high-$Q$ dielectric cavities \cite{Baranov2017}. Non-radiative decay of the excited state $S_2\rightarrow S_1$ by intramolecular vibrational relaxation (IVR) is also taken into account, with decay lifetime $1/\Gamma_{\rm IVR}\sim 0.1$ ps \cite{Hamm1997}, as well as non-radiative decay $S_1\rightarrow S_0$ into the {\it trans} and {\it cis } ground states, at rates $\Gamma_{31}$ and $\Gamma_{32}$,  respectively. The homogeneous probe absorption  linewidth is defined as $\gamma=\Gamma_{31}+\Gamma_{32}$, with typical excited state lifetimes in the range $1/\gamma\sim 0.1-0.5$  ps \cite{Fujino2001,Stuart2007}. Finally, we account for pure dephasing of the coherence between the vibrational ground states $\ket{1}$ and $\ket{2}$ at the rate $\Gamma_{\rm pd}$, with typical dephasing times in range $1/\Gamma_{\rm pd}\sim 1-100$ ps \cite{Iwata2002,Xu2002,Terentis2005,Fujiwara2008}, depending on the molecular species, solvent and temperature. Having defined a master equation for the problem, we solve for the stationary reduced density matrix of the system to obtain an expression for disorder-free medium susceptibility at the probe frequency, denoted $\chi_p$, to first order in $\Omega_p$. The derivation can be found in the Supplemental Material (SM). We then obtain the disorder-averaged susceptibility $\langle\chi_p\rangle$, by integrating  $\chi_p$--both numerically and analytically-- over random frequency configurations and dipole orientations. In what follows, we use the results of this procedure to describe the photophysics of inhomogeneously broadened molecular photoswitches inside nanoscale optical cavities.  

\begin{figure}[t]
\includegraphics[width=0.50 \textwidth]{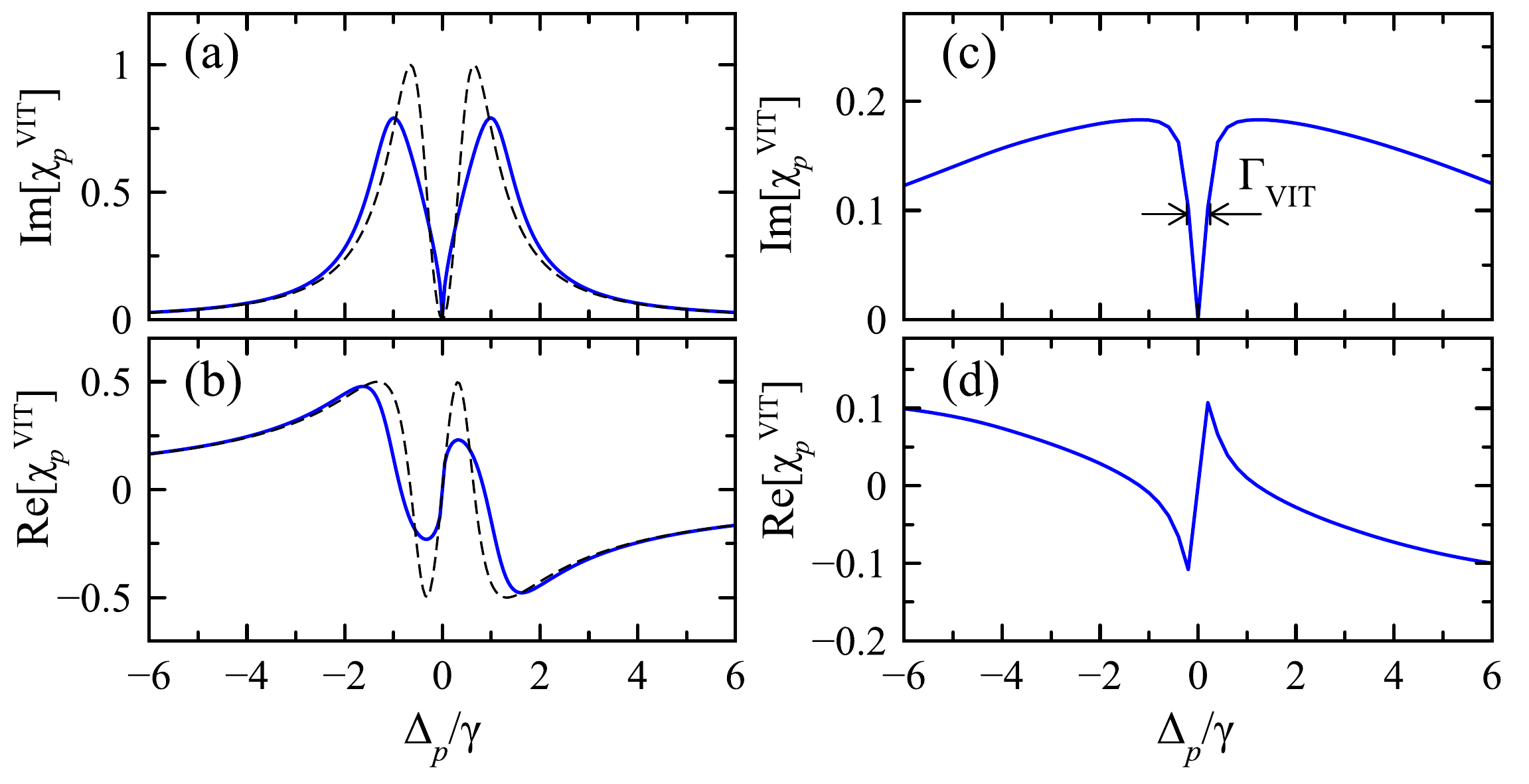}
\caption{Disorder-averaged absorptive (Im$\langle \chi_p^{\rm VIT}\rangle$) and dispersive response (Re$\langle \chi_p^{\rm VIT}\rangle$) of molecular photoswitches as a function of the probe detuning  $\Delta_p=\omega_p-\langle\omega_{31}\rangle$, for a resonant cavity field $\omega_c=\langle \omega_{32}\rangle$. (a)-(b) Uniform orientational disorder on the vacuum Rabi frequency with mean amplitude $\Omega_0=1.2\gamma$. The disorder-free lineshape is shown for comparison (dashed line). (c)-(d) Gaussian disorder on $\omega_{31}$ for $\sigma=6\gamma$ and $\Omega_c=1.2\gamma$. The VIT linewidth $\Gamma_{\rm VIT}$ is graphically defined. $\gamma$ is the homogeneous absorption linewidth.  
}
\label{fig:VIT lineshape}
\end{figure}

We begin our analysis of the intracavity response by assuming that no signal field is present ($\Omega_s=0$). In Fig. \ref{fig:VIT lineshape}, we show the corresponding disorder-averaged susceptibility $\langle \chi_p^{\rm VIT}\rangle$ as a function of the mean probe detuning $\Delta_p\equiv  \omega_p-\langle \omega_{31}\rangle$,  for the representative decay ratios $\Gamma_{\rm pd}/ \gamma=0.01$ and $\kappa/\gamma=0.0001$. We numerically average the homogeneous response over a large number of disorder configurations, separately studying orientational disorder (panels a and b) and energy disorder (panels c and d). For energy disorder, we take the random frequency fluctuations $\delta_{ji}$ from independent Gaussian distributions with the same standard deviation, i.e. $\sigma_{ji}=\sigma$. We also assume that the cavity detuning $\Delta_c\equiv \omega_c-\langle \omega_{32}\rangle$ vanishes. Figure \ref{fig:VIT lineshape} clearly shows the characteristic signatures of VIT despite disorder: strong suppression of the probe absorption (transparency) and steep dispersion (slow light), over a narrow frequency window around two-photon resonance condition $\Delta_p-\Delta_c=0$ \cite{Fleischhauer:2005}. Moreover, panels \ref{fig:VIT lineshape}a-b show that uniform orientational disorder does not  significantly alter the photoswitch response from the disorder-free case, reinforcing  previous studies on the minor importance of orientational disorder in organic cavities \cite{Litinskaya2006-disorder}, a result that can also be understood analytically (see SM for details).

Panels \ref{fig:VIT lineshape}c,d show that even for strong inhomogeneous broadening due to energy disorder ($\sigma=6\gamma$), a narrow VIT absorption dip with narrow linewidth $\Gamma_{\rm VIT}$ \cite{Gea-Banacloche:1995,Javan2002} can still persist on top of a broader Gaussian absorption background, under two-photon resonance. 
{For systems with arbitrary $\sigma$, which is the dominant energy scale in molecular photoswitches \cite{Stuart2007}, we can achieve VIT} as long as the frequency hierarchy  $\kappa\sim \Gamma_{\rm pd}\ll \Omega_c\lesssim\gamma$ holds. 
{We find survival of VIT to inhomogeneous broadening under the assumption that the static energy fluctuations of states $\ket{\tilde 1}$ and $\ket{\tilde 2}$ are equal, i.e., $\delta_{31}=\delta_{32}$ to a  very good approximation. The quality of VIT degrades as we allow the ground isomer potential minima  to fluctuate independently. This sets a constraint on the contribution of inhomogenous broadening on the vibrational Raman frequency $\omega_{21}$ to be much smaller than the excited state linewidth $\gamma$. If $\omega_{21}$ were to fluctuate significantly, the intracavity Raman dark state $\ket{D}=\sqrt{a}\ket{\tilde 1}+\sqrt{1-a}\ket{\tilde 2}$ ($0<a<1$), generated under VIT conditions \cite{Fleischhauer:2005}, would rapidly dephase, breaking the destructive interference effect that causes the VIT absorption dip.} 
However, given that both {\it cis} and {\it trans} isomers belong to the same electronic potential, it can be expected that $|\delta_{31}-\delta_{32}|/\gamma\ll 1$, independent of the molecular solvent. 

Having established the physical conditions for achieving VIT in disordered molecular photoswitches, we now discuss a cross-Kerr scheme involving the simultaneous driving of the intracavity medium by both probe and signal fields. We focus on the phase shift $\Phi$ experienced by the probe field  due to co-propagation with a signal field over a path length $L$. Attenuation at the probe frequency is quantified by the absorption coefficient $\alpha$. Under conditions of VIT, both $\Phi$ and $\alpha$ not only depend on the probe and signal field frequencies and Rabi frequencies, but also on the cavity vacuum parameters \cite{Field1993,Rice1996}. Following Ref. \cite{Kang2003}, we define the figure-of-merit $\eta \equiv \Phi/ \alpha L= {\rm Re}\langle \chi_p\rangle /2{\rm Im}\langle \chi_p\rangle $ to quantify the degree by which it is possible to interferometrically resolve the nonlinear phase shift $\Phi$ at frequency $\omega_p$. Detectable nonlinear signals require $\eta>1$. 

We developed an analytical approximation for the disorder-averaged  susceptibility $\langle \chi_p\rangle $ in the presence of both probe and signal fields.  We integrate the disorder-free  susceptibility over the random fluctuations ($\delta_{31}$, $\delta_{42}$) {assuming that they belong to independent Lorentzian distributions with different widths. We use Lorentzians instead of more realistic Gaussian distributions \cite{Biswas2002,Stuart2007} because the latter becomes analytically intractable for the cross-phase modulation signal that we seek to describe.  Our approach is motivated by previous work on Doppler-broadened EIT \cite{Gea-Banacloche:1995,Javan2002}, and is justified below by comparison with a Gaussian numerical averaging. We show in the SM, that the mean susceptibility as a function of the probe detuning $x \equiv \Delta_p$ can be written in the form}
\begin{equation}\label{eq:eta final}
2\langle\eta_p(x)\rangle = - \frac{\displaystyle x A_s(x)- \Omega_c^2 (x - x_s)}{\displaystyle \Sigma_{31} A_s(x)+ {\Omega_c^2 (\gamma_{21} + \gamma_s)} },
\end{equation}
where $\gamma_{21}\equiv (\kappa/2+\Gamma_{\rm pd})$ is the decay rate of the coherence between dressed states $\ket{\tilde 1}$ and $\ket{\tilde 2}$. \mbox{$\Sigma_{31}\equiv (\gamma/2 +\sigma_{31})$} is the total probe {\it trans} absorption linewidth, with $\sigma_{31}$ being the inhomogeneous contribution. {We have  defined the function $A_s(x)\equiv (x-x_s)^2 + (\gamma_{21} + \gamma_s)^2$, characterized by the shift $x_s \equiv \lambda_s \Delta_s$ and the width $\gamma_s\equiv \lambda_s \Sigma_{41}$,  written in terms of the dimensionless signal parameter }
\begin{equation}
\lambda_s\equiv  {\Omega_s^2 }/{(\Delta_s^2 + \Sigma_{41}^2)}, 
\end{equation}
with $\Delta_s=\omega_s-\langle \omega_{42}\rangle$ being the mean signal detuning and $\Sigma_{41}\equiv \kappa/2+\Gamma_{\rm IVR}/2 + \sigma_{42}$, where $\sigma_{42}$ is the inhomogeneous linewidth of the  {\it cis} absorption band.  {Equation (\ref{eq:eta final}) shows that within our Lorentzian disorder model, the inhomogeneous linewidths $\sigma_{31}$ and $\sigma_{42}$ simply add up to the corresponding homogeneous linewidths, allowing for a transparent interpretation of the nonlinear system photophysics.}

{Outside the cavity ($\Omega_c=0$), Eq. (\ref{eq:eta final}) simply reduces to the  linear response result $\langle \eta_p\rangle = -\Delta_p/2\Sigma_{31}$, which near  resonance gives $\langle \eta_p\rangle\ll 1$ due to the absorption of the probe. In the presence of the cavity vacuum, but without signal driving ($\Omega_s=0$), the figure-of-merit $\langle \eta_p(\Omega_s=0)\rangle \equiv \langle \eta_{\rm VIT}\rangle$ can be considered as the degree of coherence in the response of the medium to the probe field.  We show in the SM that in this VIT-regime, Eq. (\ref{eq:eta final}) reduces to 
\begin{equation}\label{eq:eta VIT function}
\langle \eta_{\rm VIT}(x)\rangle\approx \frac{\Omega_c^2 x }{ 2(\Sigma_{31} x^2 + \gamma_{21} \Omega_c^2)}\gg 1, 
\end{equation}
under the conditions $\gamma_{21}\ll x\ll \Omega_c$.  The large figure-of-merit predicted by Eq. (\ref{eq:eta VIT function}) is a direct consequence of the transparency established for the probe field by the cavity vacuum. }

For a system dominated by inhomogeneous broadening $(\gamma\ll \sigma_{31})$ inside a high-$Q$ microcavity $(\kappa\ll \Gamma_{\rm pd})$,  Eq. (\ref{eq:eta VIT function}) has a maximum at the probe detuning $x_\ast= \Omega_c \sqrt{\Gamma_{\rm pd}/\sigma_{31}}$, blue-shifted from the VIT absorption minimum, giving the optimal figure-of-merit
\begin{equation}\label{eq:eta vit}
\langle \eta^{\rm max}_{\rm VIT}\rangle = {\Omega_c}/{\sqrt{16\,\Gamma_{\rm pd}\sigma_{31}}}.
\end{equation}
With typical photoswitch parameters  $\Gamma_{\rm pd}\sim 1$ THz and $\sigma_{31}\sim 50$ THz,  reaching $\langle\eta_{\rm VIT}^{\rm max}\rangle >1$ would require $\Omega_c\geq 120$ meV. Vacuum Rabi frequencies of this order of magnitude and above can be obtained either by strongly confining the vacuum field to picoscale dimensions  \cite{Chikkaraddy:2016aa}, or by exploiting collective Rabi coupling in larger cavities \cite{Baranov2017}.

{We next discuss the behaviour of  Eq. (\ref{eq:eta final}) under simultaneous probe and signal driving $(\Omega_s>0)$. We refer to this as the VIT-Kerr regime, where the figure-of-merit is directly related with the magnitude of a cross-phase modulation signal \cite{Kang2003}.  Coupling to the signal field tends to destroy VIT, by weakly admixing the dark state $\ket{D}$ (see above), with the fast-decaying excited dressed state $\ket{\tilde 4}$. The maximum achievable figure-of-merit $\langle\eta_{\rm max}\rangle$ subject to signal driving therefore cannot exceed the VIT bound in Eq. (\ref{eq:eta vit}), i.e., $\langle \eta_{\rm max}\rangle \leq \langle \eta^{\rm max}_{\rm VIT}\rangle$ for any $\lambda_s$, with the equality sign holding only when $\lambda_s=0$. }

 {We can use Eq. (\ref{eq:eta final}) to find the constraints on $x_s$ and $\gamma_s$ that allow the nonlinear figure-of-merit to approach the VIT limit $\langle \eta^{\rm max}_{\rm VIT}\rangle $ (see SM for details). For instance, in realistic organic systems dominated by inhomogeneous broadening with $\Sigma_{41}\approx \Sigma_{31}\approx \sigma$ and assuming $\Omega_c/|\Delta_s|\leq \sqrt{\gamma_{21}/\sigma} $, our Lorentzian disorder model predicts that it is possible to reach $\langle \eta_{\rm max}\rangle\gtrsim 1$ under a single constraint on the dimensionless signal parameter, given by 
\begin{equation}\label{eq:eta threshold}
\lambda_s \lesssim \gamma_{21}/\sigma\ll 1.
\end{equation}
Small values of $\lambda_s$ are obtained by using weak signal fields, large (blue) signal detunings, or both.}

\begin{figure}[t]
\includegraphics[width=0.50\textwidth]{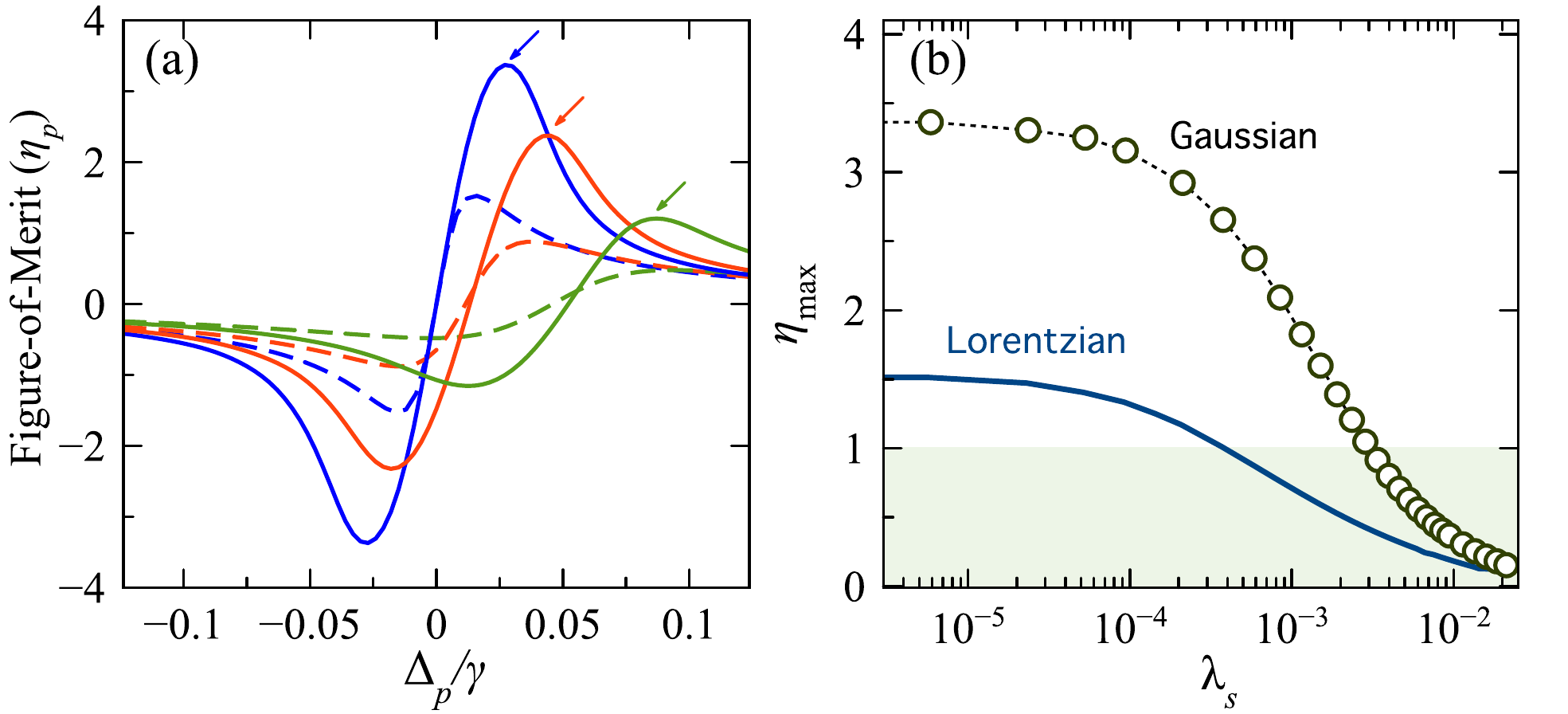}
\caption{Nonlinear figure-of-merit $\langle\eta_p\rangle$ for different signal parameters $\lambda_s$. (a) $\langle\eta_p\rangle$ as a function of probe detuning $\Delta_p$ for $\lambda_s=0$ (blue line), $\lambda_s=0.5$ (red) and $\lambda_s=1$ (green). Solid curves corresponds to a Gaussian disorder model and dashed lines to Lorentzian disorder. Arrows indicate the maximum figure-of-merit $\langle \eta_{\rm max}\rangle$ for the Gaussian case. (b)  $\langle\eta_{\rm max}\rangle$ as a function of $\lambda_s$ for Gaussian ($\sigma=5\gamma$) and Lorentzian disorder distributions with equal width (FWHM). The shaded region corresponds to signal-to-noise ratios below unity for cross-phase modulation. In both panels we set $\gamma_{21}=0.002\gamma$ and $\Omega_c = 0.8\gamma$.}
\label{fig:eta}
\end{figure}

We can now compare the predictions of our Lorentzian disorder model in Eq.  (\ref{eq:eta final}) with a more realistic Gaussian model. Figure \ref{fig:eta}a shows the $\eta(x)$-profiles predicted by the Lorentzian model and the result of numerically integrating over a large number of Gaussian configurations for $\delta_{31}$ and $\delta_{42}$, again assuming that $\delta_{31}=\delta_{32}$ as in Fig. \ref{fig:VIT lineshape}. We find that even when the Gaussian and Lorentzian noise distributions have equal widths (FWHM), Gaussian disorder consistently allows for higher values of $\langle \eta_{\rm max}\rangle $. This behaviour is captured in panel \ref{fig:eta}b, which clearly shows that for our choice of material linewidths, a system with Gaussian disorder can surpass the limit $\langle \eta_{\rm max}\rangle=1$ for values of $\lambda_s$ that are nearly an order of magnitude higher than those predicted by Eq. (\ref{eq:eta threshold}), which reduces the experimental constraints on the signal field. {Figure \ref{fig:eta}b also shows that the Gaussian asymptote on $\langle \eta_{\rm max}\rangle$ as $\lambda_s\to 0$, is about two times higher than the Lorentzian bound from Eq. (\ref{eq:eta vit}). We also find that for values of $\lambda_s$ of order $10^{-2}$ and above, the predictions of the Gaussian and Lorentzian disorder models become indistinguishable.}

Finally, we note that the single-emitter Hamiltonian $\hat H_S$  in Eq. (\ref{eq:Hamiltonian}) can be generalized to the many-particle case.  In the regime where $\kappa<\sqrt{N}\Omega_c< \gamma$, we can  replace one-body terms of the form $(\Omega_c \ket{3_i}\bra{2_i}\hat a +{\rm H.c.})$ with the collective coupling term  $(\sqrt{N}\Omega_c\ket{\alpha_{\it cis}}\bra{G_{\it cis}}\hat a +{\rm H.c.})$, where  $\ket{\alpha_{\it cis}}=\sum_{i=1}^N\ket{2_1,2_2,\cdots,3_i,\cdots,2_N}/\sqrt{N}$
is a totally-symmetric collective excitation \cite{Herrera2017-review}, and $\ket{G_{\it cis}}\equiv\ket{2_1, 2_2,\cdots, 2_N}$ is the {\it cis} ground state of the ensemble. The signal field thus weakly drives the coherence between the {\it cis} dressed ground state $\ket{G_{\it cis}}\ket{1_c}$ and state $\ket{ 4_i}\ket{1_c}$ on the $i$-th molecule. Our single-emitter results for $\langle \chi_p\rangle$ thus remain valid after replacing $\Omega_c$ by $\sqrt{N}\Omega_c$. 

In summary, we propose an organic cavity scheme to achieve large cross-phase modulation signals using {\it arbitrarily weak} probe and signal fields. The scheme exploits a long-lived vibrational coherence between the {\it cis} and {\it trans} ground vibrational states in a class of chromophores known as molecular photoswitches, to establish conditions for vacuum-induced transparency inside high-$Q$ optical microcavities. The predicted Kerr nonlinearity is found to exceed the corresponding cavity-free values by orders of magnitude, even in the presence of strong energy and orientational disorder in the organic medium. 
Our results may thus pave the way for the development of novel integrated nanophotonic devices for all-optical switching with narrow laser pulses at low power levels, using organic instead of inorganic materials \cite{Moebius:16} as the coherently driven nonlinear medium. 

\acknowledgements
We thank St\'{e}phane K\'{e}na-Cohen and Yaroslav Ispolatov for comments. M.L. also thanks the Department of Physics at USACH for hospitality during early stages of this work. 
F.H. is supported by PAI 79140030,  FONDECYT Iniciaci\'{o}n 11140158, Proyectos Basal USA 1555-VRIDEI 041731, and \mbox{Millenium Institute for Research in Optics}.

\bibliographystyle{unsrt}
\bibliography{cavityXPM}

\appendix

\counterwithin{figure}{section}

\begin{widetext}

\section{Derivation of the disorder-free susceptibility for organic photoswitches}

\subsection{Optical Bloch Equations for Cavity-Dressed States}
We model the dynamics of the molecule-cavity reduced density matrix $\hat \rho_S$ by a  quantum master equation of the form ($\hbar = 1$ throughout)
\begin{equation}\label{app:qme}
\frac{d}{dt}\hat \rho = -i[\hat H_S,\hat\rho_S] +\mathcal{L}_\kappa[\hat\rho]+\mathcal{L}_{S_2}[\hat\rho_S]+\mathcal{L}_{S_1}[\hat\rho_S]+\mathcal{D}_{\rm pd}[\hat\rho_S]
\end{equation}
where the Hamiltonian $\hat H_S$ is given by Eq. (\ref{eq:Hamiltonian}) in the main text. The first dissipator in Eq. (\ref{app:qme}) corresponds to cavity decay due to photon leakage into the far field at the rate $\kappa$,  given by the Lindblad form
\begin{equation}\label{app:cavity decay}
\mathcal{L}_\kappa[\hat \rho_S] =(\kappa/2)\left(2\hat a\hat \rho_S\hat a^\dagger - \hat a^\dagger \hat a \hat \rho_S -\rho_s\hat a^\dagger \hat a\right),
\end{equation}
The second dissipator represents the decay of the excited potential $S_2$ to $S_1$ via intramolecular vibrational relaxation (IVR) at the rate $\Gamma_{\rm IVR}$, given by
\begin{equation}\label{app:S2 decay}
\mathcal{L}_{\rm S_2}[\hat \rho_S]=(\Gamma_{\rm IVR}/2)\left(2\ket{3}\bra{4}\hat \rho_S\ket{4}\bra{3} - \ket{4}\bra{4}\hat \rho_S - \hat \rho_S  \ket{4}\bra{4}\right)
\end{equation}
The third dissipator represents non-radiative decay through a conical intersection followed by IVR, from the excited potential $S_1$ to the {\it cis} and {\it trans} manifolds in the ground potential $S_0$,  given by
\begin{eqnarray}\label{app:S1 decay}
\mathcal{L}_{\rm S_1}[\hat \rho_S]&=&(\Gamma_{31}/2)\left(2\ket{1}\bra{3}\hat \rho_S\ket{3}\bra{1} - \ket{3}\bra{3} \hat \rho_S-\hat \rho_S \ket{3} \bra{3}\right)\nonumber\\
&&+(\Gamma_{32}/2)(2\ket{2}\bra{3}\hat \rho_S\ket{3}\bra{2} - \ket{3}\bra{3} \hat \rho_S -\hat \rho_S\ket{3}\bra{3})
\end{eqnarray}
We finally introduce an {\it ad-hoc} non-Lindblad term in Eq. (\ref{app:qme}) to describe pure dephasing of the vibrational coherence between ground vibrational states in the {\it trans} and {\it cis} manifolds of $S_0$ at the rate $\Gamma_{\rm pd}$, given by
\begin{equation}\label{app:pure dephasing}
\mathcal{D}_{\rm pd}[\hat \rho_S] = -\Gamma_{\rm pd}\left(\ket{1}\bra{1}\hat \rho_S\ket{2}\bra{2}+\ket{2}\bra{2}\hat \rho_S\ket{1}\bra{1}\right)
\end{equation}
We neglect any contribution from the {\it cis}-to-{\it trans} thermal isomerization reaction in the master equation, given that the isomerization rate $\Gamma_{21}$ is negligibly low in comparison with all other dynamical processes in the problem.

Before proceeding with the  derivation of the probe susceptibility $\chi_p$ from the quantum master equation [Eq. (\ref{app:qme})], we introduce a convenient notation for the matrix elements of $\hat \rho_S$ that reads
\begin{equation}
\rho_{ij}^{mn}(t) = \bra{i; m_c}\hat \rho(t)\ket{j; n_c},
\end{equation}
where $\ket{i}$ and $\ket{j}$ represent molecular states ($i,j=1, 2,3, 4$). $\ket{m_c}$ and $\ket{n_c}$ represent cavity Fock states with photon numbers $m_c$ and $n_c$, respectively. We also define slowly-varying amplitudes for selected elements of reduced density matrix, to remove fast oscillations from the equations of motion. For the one-photon coherence $\rho_{13}^{00}(t)$, we define the slowly-varying amplitude $\sigma_{13}^{00}(t)$ by the relation $\sigma_{13}^{00}= \rho_{13}^{00} \,{\rm e}^{-i\omega_p t}$. We also define the slowly-varying variables $\sigma_{12}^{01}={\rm e}^{-i\omega_{p}t}\rho_{12}^{01}$,  $\sigma_{32}^{01}=\rho_{32}^{01}$, $\sigma_{14}^{01}={\rm e}^{-i(\omega_{p}+\omega_s)t}\rho_{14}^{01}$, $\sigma_{34}^{01}={\rm e}^{-i\omega_{s}t}\rho_{34}^{01}$, and $\sigma_{24}^{11}={\rm e}^{-i\omega_{s}t}\rho_{24}^{11}$. In terms of these slowly-varying amplitudes, we obtain from Eq. (\ref{app:qme}) the following coherence equations of motion
\begin{subequations}
\begin{eqnarray}
\dot{\sigma}_{13}^{00} &=& i(\omega_{31}-\omega_p)\,\sigma_{13}^{00} -\gamma_{31}\,\sigma_{13}^{00}-i\Omega_{\rm p}(\sigma_{33}^{00}-\sigma_{11}^{00})+i\Omega_c\sigma_{12}^{01}\;\;\;  \\
\dot{\sigma}_{12}^{01} &=& i(\omega_{21}+\omega_c-\omega_p)\sigma_{12}^{01}-\gamma_{21}\sigma_{12}^{01}-i\Omega_p\sigma_{32}^{01}+i\Omega_c\sigma_{13}^{00}+i\Omega_s\sigma_{14}^{01}\\
\dot{\sigma}_{32}^{01}&=&-i(\omega_{32}-\omega_c)\sigma_{32}^{01}-\gamma_{32}\sigma_{32}^{01}-i\Omega_c(\sigma_{22}^{11}-\sigma_{33}^{00})-i\Omega_{p}\sigma_{12}^{01}+i\Omega_s \sigma_{34}^{01}\\
\dot{\sigma}_{14}^{01}&=&i(\omega_{41}+\omega_c-\omega_p-\omega_s)\sigma_{14}^{01}- \gamma_{41} \sigma_{14}^{01} -i\Omega_p\sigma_{34}^{01}+i\Omega_s\sigma_{12}^{01}\\
\dot{\sigma}_{34}^{01}&=&i(\omega_{43}+\omega_c-\omega_s)\sigma_{34}^{01}-\gamma_{43} \sigma_{34}^{01} -i\Omega_p \sigma_{14}^{01} - i\Omega_c\sigma_{24}^{11}+i\Omega_s\sigma_{32}^{01} \\
\dot{\sigma}_{24}^{11}&=& i(\omega_{42}-\omega_s)\sigma_{24}^{11}-\gamma_{42} \sigma_{24}^{11}-i\Omega_c\sigma_{34}^{01}+i\Omega_s (\sigma_{22}^{11}-\sigma_{44}^{11})
\end{eqnarray}
\label{app:EOM}
\end{subequations}
where we introduce the decay rates 
\begin{subequations}
\begin{eqnarray}
\gamma_{31}&=&\Gamma_{31}/2+\Gamma_{32}/2\equiv \gamma/2\\
\gamma_{21}&=&\kappa/2 +\Gamma_{\rm pd}\\
\gamma_{32} &=& \kappa/2+\Gamma_{31}/2+\Gamma_{32}/2\\
\gamma_{43}&=&\kappa/2+\Gamma_{\rm IVR}/2\\
\gamma_{42}&=&\kappa+\gamma_{\rm IVR}/2\\
\gamma_{41}&=&\gamma_{43}.
\end{eqnarray}
\label{app:decay rates}
\end{subequations}
Equations (\ref{app:decay rates}a) and (\ref{app:decay rates}b) define the homogeneous probe and Raman linewidths $\gamma$ and $\gamma_{21}$, in the notation from the main text. The vibrational pure dephasing rate $\Gamma_{\rm pd}$ determines the cavity-dressed Raman lifetime $1/\gamma_{21}$ for high-$Q$ cavities with  $\kappa\ll \gamma$.

In deriving Eqs. (\ref{app:EOM}a)-(\ref{app:EOM}f), we neglect the contribution of states such as $\ket{2, 0_c}$, $\ket{3, 1_c}$, or  $\ket{4, 0_c}$, which are neither populated nor driven under our imposed assumptions of stationarity and weak signal and probe driving. Accounting for such states would result, for example, in the addition of an extra term proportional to $\sigma_{13}^{11}$ in the right-hand side of equation (\ref{app:EOM}a) for $\dot{\sigma}_{13}^{00}$, term that can be shown to vanish in the stationary limit. In other words, the set of Eqs. (\ref{app:EOM}a-f) do not correspond to a complete description of the system coherences, but can be considered as a minimal set of equations of motion that can account for the non-linear optical response of our system of interest. 

\subsection{Homogeneously-broadened intracavity susceptibility}
We look for the stationary  one-photon probe coherence $\sigma_{13}^{00}$ from Eqs. (\ref{app:EOM}). The medium polarization at the probe frequency is then given by $P(\omega_p) = d_{13}\sigma_{31}^{00}$, from semiclassical closure. Using $\Omega_p = -d_{31}E_p/2\hbar$ and $P(\omega_p) = \chi_p E_p$, the probe susceptibility $\chi_p$ for homogeneously-broadened organic molecular photoswitches can be written as
\begin{equation}\label{app:chi homogeneous}
\chi_p =  \frac{K}{(\Delta_{31}+I^{'}_0)+i(\gamma_{31}+I_0^{''})},
\end{equation}
where $K=-|d_{13}^2|/2\hbar$ is proportional to the oscillator strength of the cavity-free probe absorption peak, $\Delta_{31} =\omega_{p}-\omega_{31}$ is the probe detuning from the {\it trans} transition. In the main text we use the notation $\Delta_{31}\equiv \Delta_{p}$. 

In Eq. (\ref{app:chi homogeneous}), we have introduced the complex nonlinear quantity $I_0 \equiv I^{'}_0+iI^{''}_0$, given by
\begin{equation}\label{app:I0-1}
I_{0} =  \frac{\Omega_c^2(\Delta_{41}+i\gamma_{41})(i\gamma_{32}-\Delta_{32})}{(i\gamma_{32}-\Delta_{32})\Omega_s^2+( \Delta_{41}+i\gamma_{41}  )\Omega_p^2-(\Delta_{21}+i\gamma_{21})(\Delta_{41}+i\gamma_{41})(i\gamma_{32}-\Delta_{32})},
\end{equation}
where $\Delta_{32}=\omega_c-\omega_{32}$ is the cavity detuning from the {\it cis} absorption resonance, $\Delta_{21}=\omega_p-\omega_c-\omega_{21}$ is the two-photon (Raman) detuning, and $\Delta_{41}=\omega_p-\omega_c+\omega_s-\omega_{41}$ is the three-photon detuning. The term in the denominator proportional to $\Omega_p^2$ describes self-induced transparency, and is negligibly small in low-$\Omega_p$ limit. In what follows we use a simplified formula
\begin{equation}\label{app:I0}
I_{0} =  \frac{\Omega_c^2(\Delta_{41}+i\gamma_{41})}{\Omega_s^2 - (\Delta_{21}+i\gamma_{21})(\Delta_{41}+i\gamma_{41})}.
\end{equation}

By setting $\Omega_s=0$ in Eq. (\ref{app:I0}), the susceptibility $\chi_p$ in Eq. (\ref{app:chi homogeneous}) reduces to the standard VIT form \cite{Rice1996,Field1993} %
\begin{equation}\label{app:VIT chi}
\chi_p^{\rm VIT} = \frac{K(\Delta_{21}+i\gamma_{21})}{(\Delta_{31}+i\gamma_{31})(\Delta_{21}+i\gamma_{21})-\Omega_c^2  }.
\end{equation}

For later convenience, we introduce the dimensionless parameter 
\begin{equation}\label{app:lambda c}
\lambda_c = \frac{\Omega_c^2}{ (\Delta_{21}^2 + \gamma_{21}^2)},
\end{equation}
to rewrite Eq. (\ref{app:VIT chi}) as 
\begin{equation}\label{app:chi VIT}
\chi_p^{\rm VIT} = \frac{K}{(\Delta_{31} - \lambda_c \Delta_{21}) +i(\gamma_{31} + \lambda_c \gamma_{21})}.
\end{equation}

\section{Derivation of $\langle\chi_p^{\rm VIT}\rangle$ for disordered photoswitches without signal driving}
In this Appendix, we derive expressions for the intracavity mean probe susceptibility $\langle\chi_p^{\rm VIT}\rangle$, separately averaged over orientational and energy disorder in the organic photoswitch medium. 
 
\subsection{VIT with orientational disorder}
For an organic system with orientational disorder, disorder average can be carried out by an integration of the form $\langle\chi_p^{\rm VIT}\rangle_\theta = \int_{-\pi/2}^{\pi/2} d\theta \,\chi_p^{\rm VIT}(\cos\theta) P(\theta)$, where $\chi_p^{\rm VIT}$ is the uniform susceptibility given by Eq. (\ref{app:VIT chi}), and $P(\theta)=1/2\pi$, with $-\pi/2\leq\theta\leq \pi/2$ is a uniform distribution for the molecular {\it cis} transition dipole moment with respect to the cavity field polarization. By writting the vacuum Rabi frequency as $\Omega_c(\theta)=\Omega_0\cos\theta$, we can express the mean susceptibility $\langle \chi_p^{\rm VIT}\rangle_\theta$ as
\begin{eqnarray}\label{app:orientational integral}
\langle \chi_p^{\rm VIT}\rangle_\theta&=& K (\Delta_{21}+i\gamma_{21})\frac{1}{2\pi}\int_0^{2\pi} \frac{d\theta}{\cos^2\theta-(Z_1+iZ_2)},\nonumber\\
&=&K (\Delta_{21}+i\gamma_{21})\frac{i \,{\rm sgn(Z_2)}}{\sqrt{(Z_1+iZ_2)(1-(Z_1+iZ_2))}}
\end{eqnarray}
where $Z_1 = (\Delta_{31}\Delta_{21}-\gamma_{31}\gamma_{21})/\Omega_0^2$ and $Z_2=( \Delta_{31}\gamma_{21}+\Delta_{21}\gamma_{31})/\Omega_0^2$. Assuming for simplicity that $\gamma_{21}=0$ and rearranging terms, we arrive at the expression
\begin{equation}\label{app:chi orientational}
\langle \chi_p^{\rm VIT}\rangle_{\theta} \equiv  -K \frac{|\Delta_{21}|}{\sqrt{2}}\left[\frac{B+i (A+\Omega)}{\Omega\sqrt{A+\Omega}}\right],
\end{equation}
where we have defined $A=\Delta_{21}\left[ \Delta_{31}(\Omega_0^2- \Delta_{21}\Delta_{31}) +\Delta_{21}\gamma_{31}^2  \right]$, $B = \Delta_{21}\gamma_{31}(\Omega_0^2- 2\Delta_{21}\Delta_{31})$, and $\Omega = \sqrt{A^2+B^2}$. 
Equation (\ref{app:chi orientational}) is exact within the assumptions involved in Eqs. (\ref{app:EOM}a)-(\ref{app:EOM}f), and has excellent agreement with the numerical integration of Eq. (\ref{app:chi homogeneous}) over a large number of angular disorder configurations, as shown in Figure \ref{app:VIT orientational}a. The comparison of the VIT linewidth $\Gamma_{\rm VIT}$ for a system with orientational disorder and the disorder-free case in panel \ref{app:VIT orientational}b demonstrates the fluctuations of the dipole orientation preserves the linear scaling $\Gamma_{\rm VIT}\sim \Omega_c$ characteristic of homogeneously-broadened EIT [1,2], but with a higher slope. 
\vspace{1cm}
\begin{figure}[h]
\includegraphics[width=0.6\textwidth]{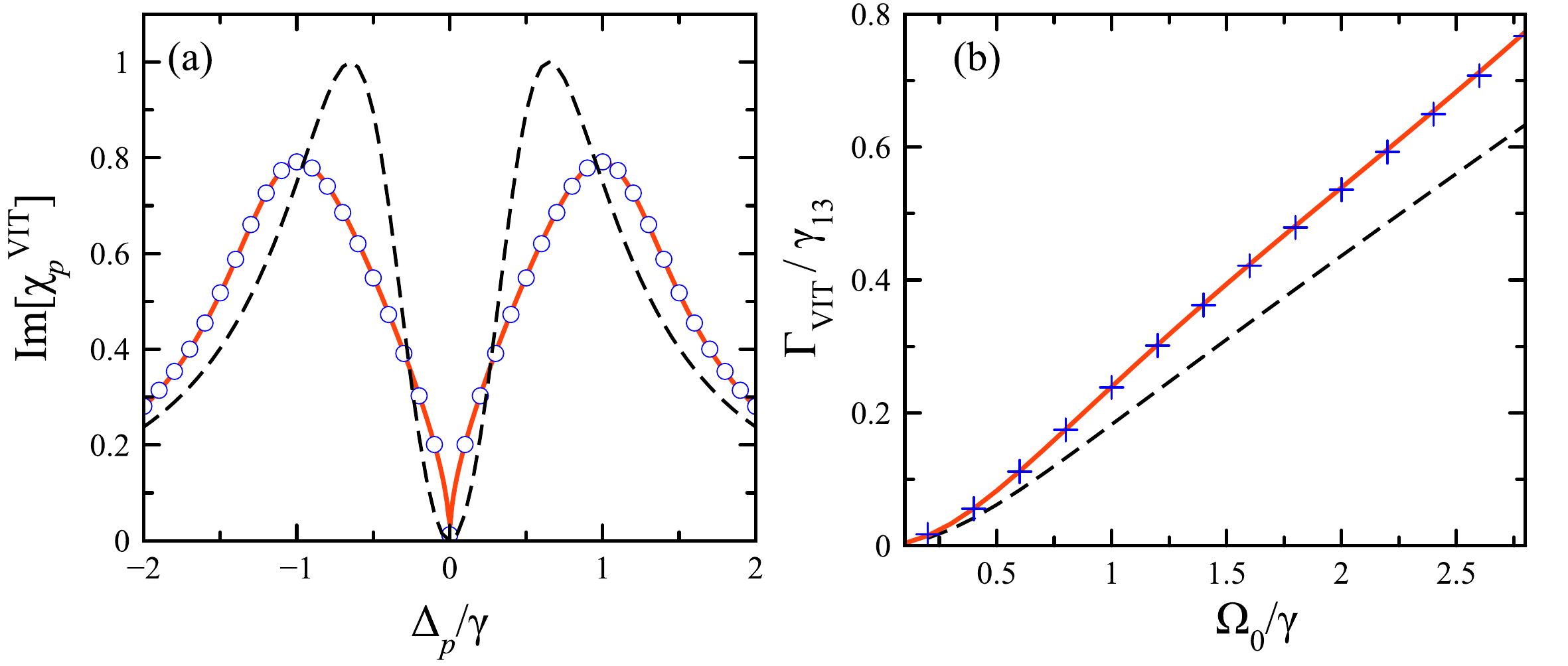}
\caption{VIT lineshape for uniform orientational disorder. (a) Average ${\rm Im}[\langle \chi_p^{\rm VIT}\rangle]$ as a function of probe detuning $\Delta_p$ for $\Omega_0=1.2\,\gamma$. Solid line corresponds to the  analytically expression in Eq. (\ref{app:chi orientational}), and circles correspond  to numerical integration over a large number of disorder configurations. The homogeneous susceptibility is also shown for comparison (dashed line). (b) Scaling of the VIT linewidth $\Gamma_{\rm VIT}$ with the Rabi amplitude $\Omega_0$, obtained both analytically (solid line) and numerically (crosses). The homogeneous scaling is also shown for comparison (dashed line).}
\label{app:VIT orientational}
\end{figure}

\subsection{VIT with energy disorder}

We average (\ref{app:chi VIT}) over energy disorder, by writing $\Delta_{31} = \langle \Delta_{31} \rangle + \delta_{31}$, and integrating over a Lorentzian distribution of the energy fluctuation given by $P(\delta_{31}) = \sigma_3/[\pi(\delta_{31}^2 + \sigma_3^2)]$, with $\sigma_3$ being the width of the energy (detuning) distribution. Note that our assumption of correlated energy fluctuations of the states $\ket{1}$ and $\ket{2}$ implies that $\Delta_{21} = \omega_p - \omega_c - \omega_{21}$ is a well-defined  quantity that does not fluctuate over the molecular ensemble. The resulting integral has one pole $(z = i\sigma_3)$ in the upper half of the complex plane. The pole arising from the denominator of $\chi_p^{\rm VIT}$ in Eq. (\ref{app:chi VIT}) always lays in the lower half-plane. We calculate the integral by closing the integration path in the upper half-plane and applying Cauchy's theorem. The result gives
\begin{equation}\label{app:chi E lambda_c}
\langle \chi_p^{\rm VIT} \rangle_E = \frac{K}{(\Delta_{31} - \lambda_c \Delta_{21}) +i(\Sigma_{31} + \lambda_c \gamma_{21})},
\end{equation}
where we have defined the combined linewidth $\Sigma_{31} = \gamma_{31} + \sigma_3$. Comparing this result with the homogeneous susceptibility in Eq. (\ref{app:chi VIT}) we see that the energy disorder average with static $\omega_{21}$ results in an additive contribution to the probe transition linewidth given by $\sigma_3$. This explains the survival of VIT conditional on the smallness of $\gamma_{21}$ and the fact that it is not sensitive to the decay rate $\gamma_{31}$, which is eliminated by the opening of the transparency window.

The analytical expression in Eq. (\ref{app:chi E lambda_c}) is in qualitative agreement with the  results in Figure \ref{fig:VIT lineshape}c,d, which were obtained numerically by  averaging Eq. (\ref{app:VIT chi}) over a large number of  detuning configurations $\delta_{31}$, distributed according to a Gaussian probability density. Small quantitative discrepancies arise due to the difference between Gaussian and Lorentzian distributions.  Gaussian distributions are more relevant for the energy disorder than Lorentzians. However,  analytically averaging Eq. (\ref{app:VIT chi}) over a Gaussian distribution results in a much less intuitive answer than our result in Eq. (\ref{app:chi E lambda_c}), in terms of integral special functions. Our Lorentzian ansatz can thus also provide a reasonably accurate description of the system response. 

\section{Derivation of $\langle \chi_p\rangle$ for  photoswitches with energy disorder}

Starting from the expression for the disorder-free susceptibility in Eq. (\ref{app:chi homogeneous}), with $I_0$ as given in Eq. (\ref{app:I0}), we introduce the random transition frequencies $\omega_{31}=\langle \omega_{31}\rangle-\delta_{31}$, $\omega_{32}=\langle\omega_{32}\rangle-\delta_{32}$, and $\omega_{42}=\langle \omega_{42}\rangle-\delta_{42}$, where $\langle \omega\rangle$ represents a mean value and $\delta$ a random static fluctuation described by the normalized distribution function $P(\delta)$. These random frequencies give rise to the random detunings $\Delta_{31}\equiv \omega_p-\langle \omega_{31}\rangle +\delta_{31}=\langle \Delta_{31}\rangle+\delta_{31}$ and $\Delta_{41}\equiv\omega_p-\omega_c+\omega_s-(\langle\omega_{42}\rangle-\delta_{42}-\langle\omega_{32}\rangle+\delta_{32}+\langle\omega_{31}\rangle-\delta_{31})=\langle \Delta_{41}\rangle +(\delta_{42}-\delta_{32}+\delta_{31})$, where $\langle \Delta_{41}\rangle$ is the mean three-photon detuning. At this point we make an assumption that the transition frequencies $\omega_{31}$ and $\omega_{32}$ undergo identical fluctuations, i.e., $\delta_{31}=\delta_{32}$, and $\delta_{42}=\delta_{41}$. We then use $\Delta_{41} \equiv \langle \Delta_{41}\rangle + \delta_{41}$, with $\Delta_{21} = \omega_p - \omega_c - \omega_{21}$ being a deterministic quantity.

We further assume that the random energy shifts $\delta_{31}$ and $\delta_{41}$ are distributed according to the Lorentzian function: $P(\delta_{31})= (1/\pi)\sigma_3/(\delta_{31}^2+\sigma_3^2)$ and $P(\delta_{41})= (1/\pi)\sigma_4/(\delta_{41}^2+\sigma_4^2)$. Under these assumptions, we can obtain the disorder-averaged susceptibility from the integral
\begin{eqnarray}\label{app:double integral}
\langle \chi_p\rangle &=& \int_{-\infty}^{\infty}d\delta_{31}\int_{-\infty}^{\infty} d\delta_{41} \left[\chi_p(\delta_{31}, \delta_{41}) P(\delta_{31}) P(\delta_{41})\right]\\
&=&K\frac{\sigma_{3} \sigma_4}{\pi^2}\int_{-\infty}^{\infty}\int_{-\infty}^{\infty} \frac{d\delta_{31}\ d\delta_{41}}{[\delta_{31}^2 + \sigma_3^2][\delta_{41}^2 + \sigma_4^2]}\times\frac{D_{21}(D_{41} + \delta_{41}) - \Omega_s^2}{D_{21}(D_{31} + \delta_{31})(D_{41} + \delta_{41}) - \Omega_s^2(D_{31} + \delta_{31}) - \Omega_c^2(D_{41} + \delta_{41})}\nonumber,
\end{eqnarray}
where we have defined the complex detunings $D_{21}\equiv \langle \Delta_{21}\rangle+i\gamma_{21}$, $D_{31}=\langle \Delta_{31}\rangle+i\gamma_{31}$ and $D_{41}=\langle \Delta_{41}\rangle+i\gamma_{41}$.  Carrying out the double contour integration in Eq. (\ref{app:double integral}), we obtain the following result:
\begin{equation}\label{app:chi E}
\langle \chi_p \rangle_E =  \frac{K}{(\Delta_{31}+I^{'}_E) + i (\Sigma_{31}+I_E^{''})},
\end{equation}
where the complex quantity $I_E = I_E^{'} + i I_E^{''}$, can be written as
\begin{equation}\label{app:IK}
I_E = \frac{\Omega_c^2 (\Delta_{41} + i \Sigma_{41})}{\Omega_s^2 - (\Delta_{21} + i \gamma_{21})(\Delta_{41} + i \Sigma_{41})}.
\end{equation}
with $\Sigma_{31} = \gamma_{31}+ \sigma_3$, and $\Sigma_{41} = \gamma_{41} + \sigma_4$, as before.  The disorder-averaged equations (\ref{app:chi E}) and (\ref{app:IK}) differ from the homogeneous equations (\ref{app:chi homogeneous}) and (\ref{app:I0}) only by the magnitude of the broadenings, which now account for the disorder.

\section{Optimal nonlinear coherent signals for photoswitches with energy disorder}
As discussed in the main text, we characterize the nonlinear signal coherence by the  figure-of-merit $\eta = {\rm Re}[\chi_p] / (2 {\rm Im}[\chi_p])$. Detectible coherent (dispersive) signals require $\eta > 1$. Our goal is to understand whether $\eta$  can still exceed unity in the presence of energy disorder for an intracavity photoswitch medium.

\subsection{Figure-of-merit in the absence of the signal field: Pure VIT regime}

In the absence of the signal field ($\Omega_s=0$), we focus on  the properties of $\eta_{\rm VIT} = {\rm Re}[\chi_p^{\rm VIT}] / (2 {\rm Im}[\chi_p^{\rm VIT}])$, with $\chi_p^{\rm VIT}$ given by the disorder-free expression in Eq. (\ref{app:chi VIT}). $\eta_{\rm VIT}$ clearly does not characterize the nonlinear phase shift of the probe in the presence of the signal field, but can still be considered to quantify the probe coherence under conditions of VIT. Extracting Re[$\chi_p^{\rm VIT}$] and Im[$\chi_p^{\rm VIT}$] from Eq.(\ref{app:chi VIT}) we obtain%
\begin{equation}\label{app:chi_VIT initial}
\eta_{\rm VIT} = - \frac{\Delta_{31} - \lambda_c \Delta_{21}}{2(\gamma_{31} + \lambda_c \gamma_{21})}.
\end{equation}
We denote the probe detuning as as $\Delta_{31}\equiv x$, assume that the cavity is on resonance with the $\ket{2} \to \ket{3}$ transition, so that $\Delta_{21} = x$, and use the definition of the cavity parameter $\lambda_c(x) = \Omega_c^2/(x^2 + \gamma_{21}^2)$ (\ref{app:lambda c}) to rewrite the equation for $\eta_{\rm VIT}$ in the form
\begin{equation}\label{app:chi_VIT initial-1}
\eta_{\rm VIT}(x) =  -\frac{x(x^2 + \gamma_{21}^2) - \Omega_c^2 x}{2[\gamma_{31}(x^2 + \gamma_{21}^2) + \gamma_{21} \Omega_c^2]}.
\end{equation}

In the absence of the cavity field ($\Omega_c = 0$), we have $\eta_{\rm free~space}(x) = -x/2\gamma_{31} \ll 1$ for $x \ll \gamma_{31}$. Inspection of Eq.~(\ref{app:chi_VIT initial-1}) reveals that the VIT enhancement, i.e. $\eta_{\rm VIT} > 1$, can be reached only when $\gamma_{21}$ is very small compared to $\gamma_{31}$, and $x$ is at most of the order of $\Omega_c \sqrt{\gamma_{21}/\gamma_{31}}$. Assuming the hierarchy of the energy scales $\gamma_{21} \ll x \ll \Omega_c, \gamma_{31}$, we can write 
\begin{equation}\label{app:eta VIT}
\eta_{\rm VIT}(x) \approx \frac{\Omega_c^2 x}{2[\gamma_{31}x^2  + \gamma_{21} \Omega_c^2]} \sim O\left( \frac{x}{\gamma_{21}} \right) \gg 1.
\end{equation}
This function has a maximum at the optimal probe detuning $x_{\ast} = \Omega_c \sqrt{\gamma_{21}/\gamma_{31}}$. Substituting $x_\ast$ into Eq. (\ref{app:eta VIT}) gives an optimal figure-of-merit for the disorder-free case of the form
\begin{equation}\label{app:eta VIT final}
\eta^{\rm max}_{\rm VIT} = \frac{\Omega_c}{4\sqrt{\gamma_{21}\gamma_{31}}}.
\end{equation}

The {\it disorder averaged} optimal figure-of-merit can be obtained within our Lorentzian disorder ansatz by simply replacing $\gamma_{31}$ with $\Sigma_{31}\equiv \gamma_{31} + \sigma_3$ from Eq. (\ref{app:eta VIT final}), to read
\begin{equation}\label{app:eta VIT average}
\langle \eta^{\rm max}_{\rm VIT}\rangle = \frac{\Omega_c}{4\sqrt{\gamma_{21}\Sigma_{31}}}.
\end{equation}
\subsection{Figure-of-merit in the presence of the signal field: VIT-Kerr regime}

Starting with the disorder-free susceptibility $\chi_p$ in Eq. (\ref{app:chi homogeneous}) and $I_0$ in Eq. (\ref{app:I0}), we can write the figure-of-merit $\eta$ as
\begin{equation}\label{app:eta full}
\eta = - \frac{\displaystyle \Delta_{31} - \frac{\Omega_c^2 (\Delta_{21} - \lambda_s \Delta_{41})}{(\Delta_{21} - \lambda_s \Delta_{41})^2 + (\gamma_{21} + \lambda_s \gamma_{41})^2}}{\displaystyle 2\left[ \gamma_{31} + \frac{\Omega_c^2 (\gamma_{21} + \lambda_s \gamma_{41})}{(\Delta_{21} - \lambda_s \Delta_{41})^2 + (\gamma_{21} + \lambda_s \gamma_{41})^2}\right] },
\end{equation}
where we have defined the dimensionless signal parameter $\lambda_s = \Omega_s^2/(\Delta_{41}^2 + \gamma_{41}^2)$. As required for consistency,  Eq. (\ref{app:eta full}) reduces to $\eta_{\rm VIT}$ in Eq. (\ref{app:chi_VIT initial}),  in the limit $\lambda_s \to 0$.

In order to understand the scaling of the {\it disorder-averaged} figure-of-merit $\langle \eta\rangle$ with the characteristic variables of the signal field ($\Omega_s,\Delta_s$), we first rewrite Eq. (\ref{app:eta full}) using $x \equiv \Delta_{31}$, $\Delta_{21} = x$, and $\Delta_{41} = x + \Delta_s$, where $\Delta_s = \omega_s- \langle \omega_{42} \rangle$.  Assuming that $|x|\ll |\Delta_s|$. We define the dimensionless $x$-independent signal parameter
\begin{equation}\label{app:lambda_s}
\lambda_s = \frac{\Omega_s^2}{\Delta_s^2 + \gamma_{41}^2}.
\end{equation}
Finally, within our Lorentzian disorder model, we  make the substitutions $\gamma_{31} \to \Sigma_{31} \equiv \gamma_{31} + \sigma_3$ and $\gamma_{41} \to \Sigma_{41} \equiv \gamma_{41} + \sigma_4$ in Eq. (\ref{app:eta full}) to obtain a mean figure-of-merit in the form
\begin{equation}\label{app:eta final}
\langle\eta(x)\rangle = - \frac{\displaystyle x [(x-x_s)^2 + (\gamma_{21} + \gamma_s)^2] - \Omega_c^2 (x - x_s)}{\displaystyle 2\left\{ \Sigma_{31}[(x-x_s)^2 + (\gamma_{21} + \gamma_s)^2] + (\gamma_{21} + \gamma_s) \Omega_c^2 \right\} },
\end{equation}
where we have introduced the shift parameter $x_s$ and the linewidth parameter $\gamma_s$, defined as
\begin{equation}
x_s \equiv \lambda_s \Delta_s = \frac{\Omega_s^2 \Delta_s}{\Delta_s^2 + \Sigma_{41}^2}
\hskip 1cm\text{ and} \hskip 1cm
\gamma_s \equiv \lambda_s \Sigma_{41} = \frac{\Omega_s^2 \Sigma_{41}}{\Delta_s^2 + \Sigma_{41}^2}.
\end{equation}

The comparison of Eq. (\ref{app:eta final}) with Eq. (\ref{app:chi_VIT initial-1}) reveals that in the presence of the signal field, the figure-of-merit $\langle \eta\rangle$ has the same functional form as in the case where no signal field is present (pure VIT). The net effect of the signal field is to blue-shift the optimal probe frequency by $x_s$ (for $\Delta_s\geq 0$) and broaden the $\eta(x)$-profile by $\gamma_s$. We can thus follow the same analysis as before, to show that the
mean figure-of-merit $\langle \eta (x)\rangle$ in the presence of the signal field, is  upper bounded by $\langle \eta^{\rm max}_{\rm VIT}\rangle = \Omega_c / (4\sqrt{\gamma_{21} \Sigma_{31}})$.

\begin{figure}[t]
\includegraphics[width=0.50\textwidth]{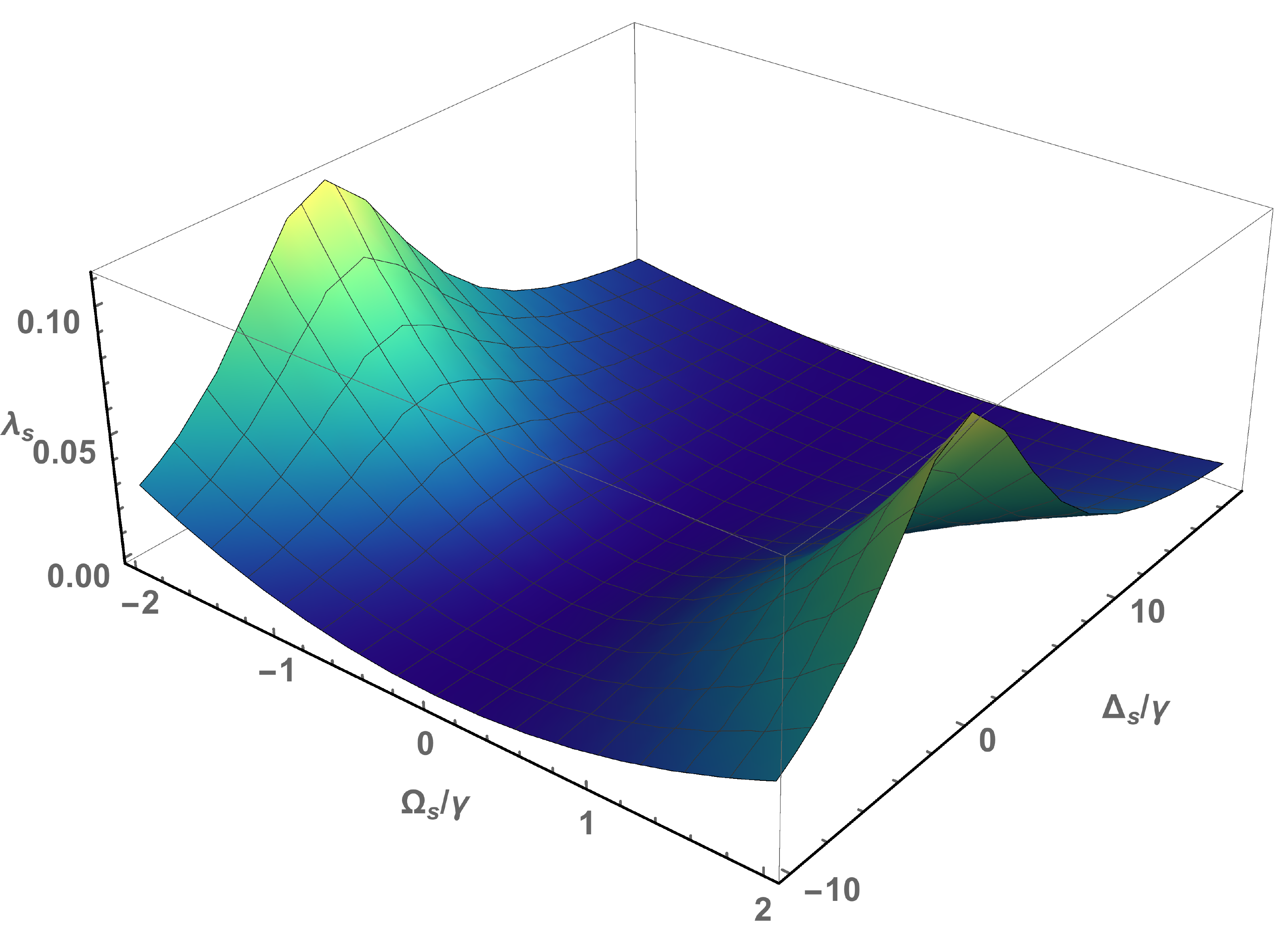}
\caption{Signal parameter $\lambda_s$ as a function of signal detuning $\Delta_s/\gamma$ and Rabi frequency $\Omega_s/\gamma$, for $\Sigma_{41}=6\gamma$.}
\label{app:lambda_s-diagram}
\end{figure}

Finally, the similarity between the VIT and Kerr schemes shows that large values of the figure-of-merit $\langle \eta\rangle$ can be obtained under the conditions $\gamma_s \leq \gamma_{21}$ and $|x_s|\leq x_\ast\equiv \Omega_c\sqrt{\gamma_{21}/\Sigma_{31}}$, which can be summarized into the \emph{signal field} parameter bound  

\begin{equation}
\frac{\Omega_s^2 }{\Delta_s^2 + \Sigma_{41}^2} \leq \frac{\gamma_{21}}{\Sigma_{41}}, \hskip 1cm {\rm and} \hskip 1cm \frac{\Omega_s^2 |\Delta_{s}|}{\Delta_s^2 + \Sigma_{41}^2} \leq \Omega_c \sqrt{\frac{\gamma_{21}}{\Sigma_{31}}}.
\end{equation}
which reduce into the single constraint in Eq. (\ref{eq:eta threshold}) of the main text, when assuming that with $\Sigma_{41}\approx \Sigma_{31}\approx \sigma$ and $\Omega_c/|\Delta_s|\leq \sqrt{\gamma_{21}/\sigma} $.  As Fig. \ref{app:lambda_s-diagram} illustrates ($\sigma_{4}$ fixed), high VIT coherence for the Kerr nonlinearity ($\lambda_s\ll 1$) can be reached by either:

\begin{itemize}

\item{Decreasing $\Omega_s$, effectively switching the signal off. The signal field party destroys VIT, as it is not protected by the transparency window.}

\item{Increasing $\Delta_s$, making the incoherent signal field less resonant and therefore less absorptive.}

\item{Increasing $\sigma_4$, effectively distributing the signal beam among many frequencies, many of which are far-detuned from $\langle \omega_{24} \rangle$ and therefore less absorptive.} 
\end{itemize}

\vspace{1cm}
\begin{enumerate}[(1)]
\item J. Gea-Banacloche, Y.-Q. Li, S.-Z. Jin, and M. Xiao, Phys. Rev. A 51, 576, 1995.
\item A. Javan, O. Kocharovskaya, H. Lee, and M. Scully,  Physical Review A 66, 013805, 2002.
\end{enumerate}

\end{widetext}

\end{document}